\documentclass{PoS}
\PoS{PoS(HEP2005)189}
\title{Quark-Lepton Complementarity with Renormalization Effects
through Threshold Corrections}
\ShortTitle{Quark-Lepton Complementarity with Renormalization Effects through
Threshold Corrections}
\author{\speaker{Sin Kyu Kang}\\
        School of Physics, Seoul National University, Korea \\
        E-mail: \email{skkang@phya.snu.ac.kr}}
\author{C. S. Kim \\
 Department of Physics, Yonsei University, Korea \\
 E-mail: \email{cskim@yonsei.ac.kr}}
 \author{Jake Lee \\
 Department of Physics, Yonsei University, Korea \\
 E-mail: \email{jilee@cskim.yonsei.ac.kr}}

\abstract{The recent experimental measurements of the solar
neutrino mixing angle $\theta_{sol}$ and the Cabibbo mixing angle
$\theta_C$ reveal a surprising relation, $ \theta_{sol}+\theta_C
\simeq \frac{\pi}{4} $. We review that while this empirical relation 
has been interpreted as a support of the idea of grand unification, 
it may be merely  accidental in the sense that reproducing the relation
at a low energy in the framework of grand unification may depend
strongly on the renormalization effects whose size can vary with
the choice of parameter space. We note that the lepton mixing
matrix derived from quark-lepton unification can lead to a shift
of the complementarity relation at low energy. While the
renormalization group effects generally lead to additive
contribution on top of the shift, we show that the threshold
corrections which may exist in some intermediate scale new physics
such as supersymmetric standard model can diminish it, so we can
achieve the complementarity relation at a low energy.}

\FullConference{International Europhysics Conference on High
Energy
Physics\\
         July 21st - 27th 2005\\
         Lisboa, Portugal}
\def\bea{\begin{eqnarray}}
\def\eea{\end{eqnarray}}

\begin{document}
Recently, it has been noted that the solar neutrino mixing angle
$\theta_{sol}$ and the Cabibbo angle $\theta_C$ reveal a surprising
relation \bea \theta_{sol}+\theta_C \simeq \frac{\pi}{4}, \label{qlc}\eea
which is satisfied by the experimental results
$\theta_{sol}+\theta_{C}=45.4^{\circ}\pm 1.7^{\circ}$ to within a
few percent accuracy \cite{fit2}. This quark-lepton
complementarity (QLC) relation (\ref{qlc}) has been interpreted  as an
evidence for  quark-lepton
unification \cite{raidal}. Yet, it
can be a coincidence in the sense that reproducing the exact QLC
relation (\ref{qlc}) at low energy scale  in the framework of grand
unification depends on the renormalization effects whose size can
vary with the choice of parameter space \cite{kang}.

A parametrization of
the PMNS mixing matrix in terms of a small parameter whose
magnitude can be interestingly around $\sin\theta_C$ has been
proposed as follows \cite{giunti}: \bea
U_{\rm PMNS}=U^{\dagger}(\lambda)U_{\rm bimax}~. \label{framp}
\eea Here $U(\lambda)$ is a mixing matrix parameterized in terms
of a small parameter $\lambda$  and $U_{\rm bimax}$ corresponds to
the bi-maximal mixing matrix \cite{bimax}.
In this work, we show that the lepton mixing matrix
given in the form of Eq. (\ref{framp}) with
$U(\lambda)\sim U_{\rm CKM}$ can be indeed realized in the
framework of grand unification with symmetric Yukawa matrices when
we incorporate seesaw mechanism, and examine whether or not
$U_{\rm PMNS}$  given by
(\ref{framp}) can predict the QLC relation (\ref{qlc}) exactly.

It is necessary to take into account the
renormalization effects on $U_{\rm PMNS}$ when one
compares the prediction at a high energy scale with the QLC
relation observed at low energy scale \cite{rgee0}.
In MSSM with large $\tan\beta $ and the
quasi-degenerate neutrino mass spectrum, the RG effects are
generally large and can enhance the mixing angle $\theta_{12}$ at
low energy \cite{rgee0}. Such an enhancement of
$\theta_{12}$ is not suitable for achieving the QLC relation (\ref{qlc})
at low energy .
In this work, we show that  the sizeable {\it threshold
corrections} which may exist in the MSSM
\cite{chan,slepton} can diminish the deviation from the
QLC relation while keeping $\theta_{23}$ almost
maximal and $\theta_{13}$ small, so that the QLC relation at low
energy can be achieved when the RG effects are
suppressed.

The quark Yukawa matrices $Y_u,Y_d$ are given by
$Y_u=U_uY_u^{diag}V_u^{\dagger},~~Y_d=U_dY_d^{diag}V_d^{\dagger}$,
from which CKM mixing matrix is given by
$U_{\rm CKM}=U^{\dagger}_uU_d~$. 
The charged lepton Yukawa matrix is given by
$Y_l = U_lY_l^{diag}V^{\dagger}_l$.
For the neutrino sector, we introduce one right-handed singlet
neutrino per family which leads to the seesaw mechanism, according
to which the light neutrino mass matrix is given by  $M_{\nu}
= \left( U_{0}M_{\rm Dirac}^{diag}V^{\dagger}_0 \right)
\frac{1}{M_R} \left(V^{\ast}_0 M_{\rm Dirac}^{diag} U^T_0
\right),\label{lep1}$ where $U_0$ and $V_0$ are the
left-handed and right-handed mixing matrices of $M_{\rm Dirac}$, respectively.
We can then rewrite $M_{\nu}$ as follows
$M_{\nu} = U_0V_M M_{\nu}^{diag}V_M^T U_0^T,$ where $V_M$ represents the rotation of
$M_{\rm Dirac}^{diag}V^{\dagger}_0 \frac{1}{M_R}V^{\ast}_0 M_{\rm Dirac}^{diag}$~.
Then, $U_{\rm PMNS}$ is given by
\bea U_{\rm PMNS} = U^{\dagger}_l U_{\nu} =
U^{\dagger}_lU_0V_M~.\label{lep3} \eea

Now, let us consider how $U_{\rm PMNS}$ given by Eq.
(\ref{lep3}) can be related with $U_{\rm CKM}$ in the quark-lepton unification. \\
{\bf (A) Minimal quark-lepton unification :}
Since the down-type quarks and the charged leptons are in general
assigned into a multiplet in grand unification, we can assume that
$Y_e=Y^T_d,~~~Y_u=Y^T_u$.
Then, we deduce that
$U_l=V_d^{\ast}$ from which $
  U_{\rm PMNS}=V^T_d U_0 V_M $.
{}From this expression for $U_{\rm PMNS}$, we see that
the contribution of $U_{\rm CKM}$ may appear in $U_{\rm PMNS}$
when $Y_{\nu}=Y_u$ which can be realized in $SO(10)$.
Then, one can obtain
$U_{\rm PMNS}
= V^T_d U_d U_{\rm CKM}^{\dagger}V_M $ and then requiring
symmetric form of $Y_d$, we finally
obtain \bea U_{\rm PMNS}=U^{\dagger}_{\rm CKM}V_M, \label{sym}\
\eea where $V_M$ has bi-maximal mixing pattern.
In this way, $U_{\rm PMNS}$ can be connected with $U_{\rm CKM}$.
To see whether the parametrization of $U_{\rm PMNS}$ given by
(\ref{sym}) can lead to the QLC relation (\ref{qlc}), it is convenient to
present $U_{\rm PMNS}$ for the CP-conserving case as follows:
\bea
U_{\rm PMNS} =&U^{\dagger}_{\rm CKM}U_{23}^m U_{12}^m
\equiv
U_{23}(\theta_{23})U_{13}(\theta_{13})U_{12}(\frac{\pi}{4}-\theta_{12})~,
\label{ckm3} \eea where $U_{12}^m$ and $U_{23}^m$ correspond to
the maximal mixing between (1,2) and (2,3) generations,
respectively. The
solar neutrino mixing $\sin\theta_{sol}$ then becomes $\sin\theta_{sol}\simeq
\sin\left(\frac{\pi}{4}-\theta_C\right)
+\frac{\lambda}{2}(\sqrt{2}-1)$. Thus, we see
that the neutrino mixing matrix (\ref{ckm3}) originating from the
quark-lepton unification obviously leads to a shift of the
relation (\ref{qlc}). Numerically, the shift amounts to $\delta
\theta_{sol} \simeq 3^{\circ}$ and we can expect that
renormalization effects on Eq. (\ref{ckm3})
may fill the gap between the QLC relation and the prediction
for $\sin\theta_{sol}$ from high energy mixing matrix.\\
{\bf (B) Realistic quark-lepton unification :}
Although the minimal quark-lepton unification can lead to an
elegant relation between $U_{\rm PMNS}$  and $U_{\rm CKM}$ as shown
above, it indicates undesirable mass relations between quarks
and leptons at the GUT scale such as $m_d^{diag}=m_l^{diag}$.
Recently,  a desirable form of $U^{\dagger}_lU_0$ has been
suggested based on a well known empirical relation $|V_{us}|
\simeq \sqrt{\frac{m_d}{m_s}}\simeq 3\sqrt{\frac{m_e}{m_{\mu}}}$
\cite{pakvasa}, from which $\sin\theta_{sol}$ is given by $
\sin\theta_{sol} \simeq \sin\left(\frac{\pi}{4}-\theta_C\right)
+\frac{\lambda}{2}\left(\sqrt{2}-\frac{1}{3}\right)~$.
Numerically, the deviation from the QLC relation amounts to
$\delta \theta_{sol}\simeq 7^{\circ}$. We consider a possibility
that the threshold corrections can diminish the deviation from
the QLC relation.

 Now, let us examine how the renormalization effects can diminish
 the deviation from the QLC relation.  In general, the radiative
corrections to the effective neutrino mass matrix are given by:
\bea
M_{\nu} = I\cdot M^0_{\nu} \cdot I
= I\cdot U^T_{\rm CKM}U_{23}^{m^{\ast}} M_{D12}U^{m\dagger}_{
23}U_{\rm CKM} \cdot I~, \eea where  $M_D=Diag[m_1,m_2,m_3]$,
$M_{D 12}=U^{m^{\ast}}_{12} M_D U^{m\dagger}_{12}$, and the matrix
$I\equiv I_A\delta_{AB}, (A,B = e, \mu, \tau)$ stands for the
radiative corrections. The correction $I$ generally consists of
two parts $I=I^{RG}+I^{TH}$ where $I^{RG}$ are RG corrections
and $I^{TH}$ are electroweak scale threshold corrections \cite{chan}.
The typical size of RG corrections $I^{RG}$ is
known to be about $10^{-6}$ in the SM and MSSM with small
$\tan\beta$, and thus negligible. In addition, supersymmetry can
induce flavor dependent threshold corrections related with slepton
mass splitting which can dominate over the charged lepton Yukawa
corrections \cite{slepton}.  We have numerically checked that RG
evolution from the seesaw scale to the weak scale {\it enhances}
the size of $\theta_{12}$ in the case that $\theta_{13}$ and
$\theta_{23}$ are kept to be small and almost maximal mixing,
respectively. Thus, the case of sizable RG effects is not suitable
for our purpose. Instead, we examine whether the threshold
corrections can be suitable for diminishing the deviation from the
QLC relation while keeping $\theta_{23}$ nearly maximal and
$\theta_{13}$ small in the case that RG effect is negligible.
 To achieve our goal, we note that the contribution
$I_e$ should be dominant over $I_{\mu,\tau}$ because only $I_e$
can lead to the right amount of the shift of $\theta_{12}$ while
keeping the changes of $\theta_{23}$ and $\theta_{13}$ small.
Taking $|I_e| >> |I_{\mu,\tau}|$, the neutrino mass matrix
corrected by the leading contributions is  rewritten as follows:
 \bea M_{\nu}
 &\simeq& U^{T}_{\rm CKM}U_{23}^{m^{\ast}}\left[I_D+I_e
 \Lambda_{\lambda}
 \right]M_{D 12}\left[I_D+I_e\Lambda^{\dagger}_{\lambda}\right]
 U^{m\dagger}_{23} U_{\rm CKM}~, \label{corr}
 \eea
 where $I_D$ is $3\times 3$ identity matrix, and
 the matrix $\Lambda_{\lambda}$ is given in terms of $\lambda$.

To see how much the lepton mixing angles can be shifted by $I_e$,
we do numerical analysis in a model independent way based on the
form given by Eq. (\ref{corr}), and by taking $\Delta
m^2_{sol}\equiv m_2^2-m_1^2\simeq 7.1\times 10^{-5}~\mbox{eV}^2$
and $\Delta m^2_{atm}\equiv m^2_3-m_2^2\simeq 2.5\times
10^{-3}~\mbox{eV}^2$. Varying the parameter $I_e$
and the smallest light neutrino mass $m_1$, we find which
parameter set $(I_e, m_1)$ can lead to the QLC relation exactly
and the results are presented in Table I. The first and the second
row in Table I correspond to the minimal unification and
realistic case, respectively.
 In our analysis, we have also checked that
 $\theta_{23}$ is almost unchanged,
whereas  the shift of $\theta_{13}$ is about $1^\circ$ for both
cases (A) and (B). {}From Table I, we see that a larger value of
$I_e$ is required to achieve the relation (\ref{qlc}) as $m_1$ goes down.
\begin{table}
\begin{tabular}{|c|c|c|c|c|}\hline
& $m_1$ (eV) &0.15& 0.1 & 0.05 \\ \hline
 $I_e$ & (A) &$-4.0\times 10^{-5}$ & $-8.5\times 10^{-5}$ & $-3.4\times 10^{-4}$ \\
 $I_e$ & (B)   & $-1.0\times 10^{-4}$ &$-2.2\times 10^{-4}$ & $-8.6\times 10^{-4}$\\
\hline
\end{tabular}
 \caption
 {
 Parameter set $(I_e, m_1)$ leading to the QLC relation.
 The rows (A) and (B) correspond to the minimal unification and
 realistic case, respectively.
 }
\end{table}

To achieve $|I_e|>>|I_{\mu,\tau}|$, we can consider a dominant
contribution of chargino (pure W-ino) to $I_e$ in MSSM, and
it turns out that the size of $|I_e|$ is about 10 times larger than
that of $|I_{\mu,\tau}|$ for
 $M_e\sim 2M_{\mu,\tau}$, and the value of $I_e$ becomes negative and of the order
 of $10^{-4} \sim 10^{-3}$ for
 $x_e\gtrsim 0.65$, which are required to achieve the exact QLC relation at low energy.

\end{document}